\documentstyle[psfig,11pt]{article}
\begin{document}
{\hfil\hfill \bf EFI 97-15}

\vspace{3mm}
\begin{center}

{\Large \bf Limits on the isotropic diffuse flux of}

\vspace{1mm}

{\Large \bf ultrahigh energy $\gamma$-radiation}

\vspace{6mm}

M.C. Chantell, C.E. Covault, J.W. Cronin, B.E. Fick,
J.W. Fowler, 

L.F. Forston, K.D. Green, B.J. Newport, R.A. Ong,
S. Oser

{\em The Enrico Fermi Institute, The University of Chicago,
Chicago, IL 60637, USA}

\vspace{5mm}

M.A. Catanese$^\dag$, M.A.K. Glasmacher, J. Matthews,
D.F. Nitz, D. Sinclair, J.C. van der Velde

{\em Department of Physics, The University of Michigan,
Ann Arbor, MI 48109, USA}

\vspace{5mm}

D.B. Kieda

{\em Department of Physics, The University of Utah,
Salt Lake City, UT 84112, USA}

\vspace{4mm}

\centerline{(Submitted to Phys. Rev. Lett.)}

\vspace{1mm}

\end{center}

\vspace{10mm}

\centerline{\bf Abstract}

Diffuse ultrahigh energy $\gamma$-radiation
can arise from a variety of 
astrophysical sources,
including the interaction of $10^{20}\,$eV cosmic
rays with the $3\,$K microwave background radiation
or the collapse of
topological defects created in the early Universe.
We describe a sensitive search for diffuse $\gamma$-rays
at ultrahigh energies using the 
CASA-MIA experiment.
An isotropic flux of 
radiation is not detected, and we place stringent
upper limits on the fraction of the $\gamma$-ray component relative to
cosmic rays ($< 10^{-4}$) at energies from
$5.7 \times 10^{14}\,$eV to $5.5\times 10^{16}\,$eV.
This result represents the first comprehensive constraint
on the $\gamma$-ray flux at these energies.

\newpage


We have known about the existence of
ultrahigh energy (E$ > 10^{14}\,$eV) cosmic rays
for more than 50 years \cite{ref:auger}, and yet their
origin remains mysterious.
Similarly, the possibility of a $\gamma$-ray component of
the cosmic radiation remains an important
open question in astrophysics \cite{ref:turner}.
Direct detection of ultrahigh energy photons 
by satellite and balloon-borne experiments 
is difficult because of the low
fluxes.
Fortunately, detection can be accomplished by
large area 
($> 10^5\,$m$^2$) 
ground-based detectors 
using the air shower technique.
This paper reports a sensitive search for a high energy diffuse
$\gamma$-ray component
of the cosmic rays made by the Chicago Air Shower Array-
Michigan Muon Array (CASA-MIA) experiment.

There are a variety of possible sources for a diffuse radiation
at ultrahigh energies.
Conventional astrophysical scenarios include:
(a) unresolved point sources, such as active galactic nuclei (AGN)
which may be the source of extremely high
energy (E$> 10^{18}\,$eV)
cosmic rays \cite{ref:agn},
(b) decays of neutral pions produced by the collisions
of cosmic rays with interstellar gas and dust \cite{ref:galactic},
and
(c) electromagnetic cascades resulting from the interactions of
extremely high energy cosmic rays with the cosmic microwave
background radiation (CMBR) 
\cite{ref:high_end}.
In scenario (b),
the diffuse $\gamma$-ray flux is expected
to be concentrated in the galactic plane. 
We have already reported a separate search for a
diffuse galactic $\gamma$-ray component \cite{ref:covault}.

In addition to these conventional sources, there might be
new astrophysical
or cosmological phenomena giving rise to a diffuse radiation
at ultrahigh energies.
For example, there is considerable speculation that the highest
energy 
(E$> 10^{20}\,$eV)
cosmic rays 
may result from the collapse
or annihilation of topological defects
formed at grand unified energy scales in the early Universe 
\cite{ref:td}.
With such scenarios, the production of cosmic rays is accompanied by
electromagnetic cascades leading to copious $\gamma$-ray production.

In any scenario in which $\gamma$-rays are produced at cosmological
distances, we expect a substantial attenuation of the signal
as a result of
pair-production with intergalactic radiation fields.
$\gamma$-rays near $10^{15}\,$eV should interact with the CMBR with
an attenuation length $\sim 20\,$kpc \cite{ref:high_end}.
Nevertheless, cascading from higher energies in the presence
of a weak intergalactic magnetic field can propagate
$\gamma$-rays over significant distances \cite{ref:Protheroe}.
At the lower energies accessible by ground-based experiments
($10^{12} - 10^{14}\,$eV), absorption by infrared (IR) 
radiation is also expected, and may be significant,
although the exact amount of 
absorption is still very uncertain \cite{ref:ir}.

Ultrahigh energy photons and cosmic rays arrive at Earth,
interact in the atmosphere, and create extensive 
air showers (EAS).
EAS detectors sample the secondary shower particles
(primarily electrons and muons).
In showers initiated by cosmic-ray nuclei,
muons are generated predominantly from the decays of charged mesons
produced in hadronic interactions.
$\gamma$-ray showers, however, are largely electromagnetic in nature
since the cross section for meson photoproduction 
is much smaller than that for pair production,
by a factor of approximately $10^{-3}$ \cite{ref:hera}.
As a result, we expect a much smaller fraction of secondary hadrons
in $\gamma$-ray showers relative to cosmic-ray showers, and correspondingly
far fewer muons.
Simulations based on the MOCCA \cite{ref:mocca} and
CORSIKA \cite{ref:corsika} programs indicate that $\gamma$-ray
showers contain, on average, 3--4\% of the number of muons
as showers initiated by protons of the same energy.
These results, which agree with other calculations \cite{ref:mupoor},
are based on conventional electromagnetic and photoproduction
couplings. 
The possibility of an anomalously high photoproduction cross
section has been experimentally ruled out at HERA
up equivalent laboratory
energy of $\sim 2 \times 10^{13}\,$eV
\cite{ref:hera}.
The simulation cross sections are in agreement with these findings, and
are assumed to continue smoothly at higher energies.
The technique for distinguishing
$\gamma$-rays from cosmic rays
by identifying muon-poor EAS was suggested more than thirty
years ago \cite{ref:early_mupoor}, and
its main limitation results from downward
fluctuations in the muon content of hadronic showers.
In all experiments prior to CASA-MIA, the muon detectors have not been
sufficient in size to provide a clear statistical separation between the
$\gamma$-ray and cosmic-ray regimes.

Detections of ultrahigh energy $\gamma$-rays using the
muon-poor air shower technique
were reported in the 1960's
by a Polish-French collaboration \cite{ref:lodz}
and by the BASJE collaboration \cite{ref:basje}.
Later detections were reported by the Tien Shan \cite{ref:tienshan}
and Yakutsk \cite{ref:yakutsk}
experiments. 
The first significant upper limit on the diffuse photon flux at
ultrahigh energies 
was reported by the Utah-Michigan experiment using
a portion
of the eventual CASA-MIA detector \cite{ref:matthews}.
Upper limits have been reported by the
HEGRA experiment,
based on the study of the lateral distribution of Cherenkov light in EAS
\cite{ref:hegra}, and by the EAS-TOP detector
based on muon-poor air showers \cite{ref:eastop}.
The latter result used 
an analysis procedure in which the limit value 
was apparently optimized
by an arbitrary cut on shower size.
An upper limit 
based on the electromagnetic component of
EAS recorded in emulsion chambers has been claimed \cite{ref:he}, but
this result is controversial \cite{ref:hegra}.

The CASA-MIA experiment is located in Dugway, Utah, USA
($40.2^\circ\,$N, $112.8^\circ\,$W) 
and consists
of a surface array
of 1089 scintillation detector stations 
enclosing an area
of $2.3\times 10^5\,$m$^2$, and a muon array
of 1024 scintillation counters having
an active area of
$2,500\,$m$^2$ \cite{ref:nim}.
The muon array is more than ten times
larger than any other EAS muon detector built
for gamma-ray astronomy.
For this analysis, we use data taken 
between
January 1, 1992 and January 7, 1996,
when the full
experiment was operational.

For each event recorded by CASA-MIA, 
we estimate the shower size from the number of
{\em surface detector stations} struck,
the shower core from the location of maximum particle
density,
and the shower direction 
from detector timing information \cite{ref:nim}.
The number of {\em detected muons} is determined
from those muon counters hit
within a narrow window 
(width $\sim 100\,$nsec) around the expected arrival time.
On average, we expect 0.65 muons per event from accidental muon hits.

We have developed a set of data quality cuts,
described elsewhere \cite{ref:binary},
to remove periods of time
in which detector problems could potentially bias the
reconstruction procedures.
We require individual events to have valid information
from the underground muon array,
a shower core greater than $15\,$m from the physical edge of the
surface array, and a reconstructed zenith angle less than $60^\circ$.
We remove events 
in which a mismatch was likely to have 
occurred between the
information derived from the surface array and that
derived from the muon array.
The fraction of these mismatched events is $0.034$\% of the total
event sample, and is independent of shower size.
After all cuts, the final data sample consists 
$1.370\times 10^{9}$ events, which
is larger by more than an order of magnitude
than those used in other recent searches 
\cite{ref:matthews,ref:hegra,ref:eastop}.

The median $\gamma$-ray energy for EAS that trigger CASA-MIA is
$\sim 10^{14}\,$eV.
At this energy, the mean number of detected muons is $\sim 7.5$,
and 4.1\% of the events have zero detected muons.
These zero muon events result from downward statistical
fluctuations in the
number of detected muons in ordinary cosmic-ray EAS.
To greatly improve the background rejection, we
consider data samples at higher energies in which the
average number of detected muons is larger.
We select five non-independent data samples by requiring
the minimum number of stations to be greater than 50, 100, 250, 500, 
and 700.
The total number of events, $N_{\rm tot}$, and the average number
of detected muons for each sample is
shown in Table~\ref{tab:samples}.

\begin{table}[ht]
\caption{
Data samples used.
The minimum number of stations required is listed in
the first column.
The total number of events, $N_{\rm tot}$,
the average number of muons, $< N_\mu >$,
and the median comsic ray energy, $E_{cr}$ are given
in the second, third, and fourth columns, respectively.}
\label{tab:samples}
\medskip
\begin{tabular}{rccc}
Sample & $N_{\rm tot}$ & $< N_\mu >$ &$E_{cr}$ (TeV) \\
\hline
$ >  50$ stations & $6.9090 \times 10^{7}$ & 40.7  & 575    \\
$ > 100$ stations & $1.6042 \times 10^{7}$ & 78.5  & 1,350  \\
$ > 250$ stations & $1.1863 \times 10^{6}$ & 210.6 & 5,000  \\
$ > 500$ stations & $71,534$               & 470.1 & 22,000 \\ 
$ > 700$ stations & $11,572$               & 641.3 & 55,000 \\
\end{tabular}
\end{table}

We use a simulation to reproduce the observed muon number distribution
and to estimate the median cosmic ray
energy of each data sample, as shown in
Table~\ref{tab:samples}.
A library of artificial EAS are generated by
the CORSIKA simulation code \cite{ref:corsika}
using the known cosmic-ray energy spectrum
(see discussion in \cite{ref:crab}) and 
a chemical
composition that is independent of energy and consistent with
spacecraft measurements \cite{ref:swordy}.
The experimental detection efficiency is determined by observing
the deviation of the shower size spectrum from a power law.
Using the simulated muon size and assuming a Greisen lateral distribution
function \cite{ref:greisen}, the 
number of detected muons is determined
by a detector simulation which
accounts for all known parameters of the muon detector
(accidental muon hits, detection efficiency, etc.).
We vary the median cosmic-ray energy in the simulation
to obtain the best match between the simulation and data
distributions of the number of detected muons.
The simulation reproduces the main features of the
data for all samples.
There is good agreement between
the cosmic-ray energy values
estimated from the simulation and those derived using
the constant intensity method \cite{ref:crab}.

\begin{figure}[ht]
\centerline{\psfig{file=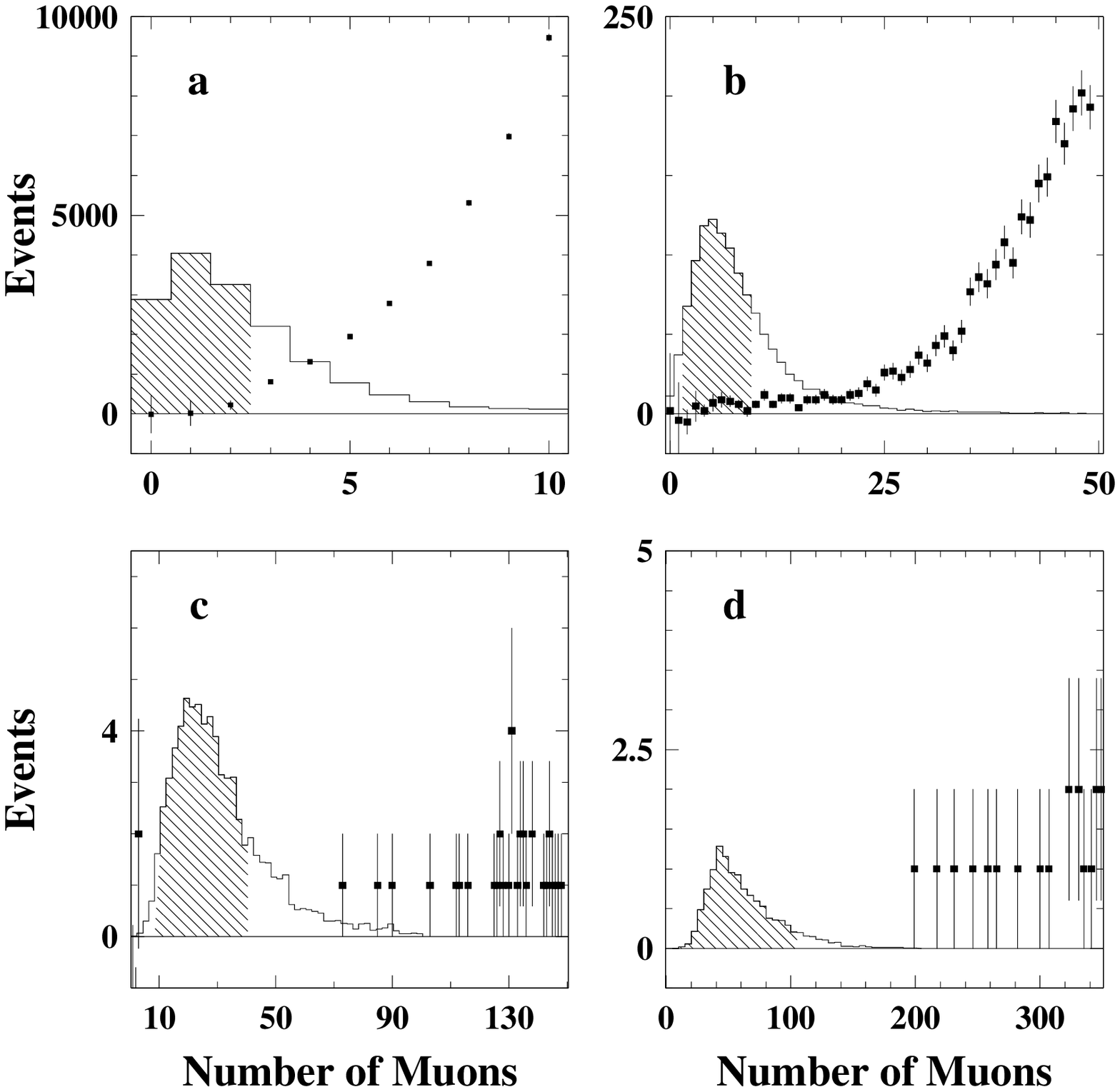,width=5.0in}}
\medskip
\caption[...]
{Histograms of the number of detected muons for four 
data samples: (a) $> 100$ stations, (b) $> 250$ stations,
(c) $> 500$ stations, and (d) $> 700$ stations.
The data are represented 
by the points with error bars.
The error bars account for the statistical uncertainty in the
numbers of events and the systematic uncertainty in the
subtraction of mismatched events.
The $\gamma$-ray simulation results are represented by the solid lines,
normalized to $10^{-3}$ of the cosmic-ray flux.
The hatched regions correspond to the 
$\gamma$-ray signal regions as described in the text.
\label{fig:low}}
\end{figure}

We use the simulation to estimate the number of muons expected
for showers initiated by $\gamma$-ray primaries.
Fig.~\ref{fig:low} shows the expected distributions 
in comparison with the data.
For each sample, we define a signal region of the muon distribution in which
the bulk of the $\gamma$-ray signal would be contained, and
in which the known backgrounds are largely excluded.
For example, in the $> 250$, $> 500$, and $> 700$ station samples,
we exclude the bins at low muon content where we know
there is potential background from mismatched events.
No significant excesses consistent with a $\gamma$-ray signal
are seen in any sample.
Therefore, we assume that all events within the signal regions are background
and set upper limits on the $\gamma$-ray flux and
on the $\gamma$-ray fraction of the cosmic rays.
We estimate, $N_{90}$, the 90\% C.L. upper limit on the
number of detected events, using standard statistical methods \cite{ref:PDB},
and use the simulation to evaluate the efficiency for $\gamma$-ray
detection, $\epsilon_\gamma$, when the muon cut is applied.
The upper limit on the fraction of the $\gamma$-ray integral flux relative to
the cosmic-ray integral flux,
$I_\gamma / I_{cr}$, is given by:

\begin{equation}
{ {I_\gamma}\over{I_{cr}} } \ < \ 
{ {N_{90}} \over {\epsilon_\gamma N_{\rm tot}} } 
\Biggl(
{ {E_{cr}} \over {E_\gamma} }
\Biggr)^{\alpha} 
\ \ ,
\label{eq:fract}
\end{equation}

\noindent where 
$E_{cr}$ is the median cosmic-ray energy,
$E_{\gamma}$ is the median $\gamma$-ray energy,
and $\alpha$ is the integral cosmic-ray spectral index 
($\alpha = -1.7$, E$ < 3\times 10^{15}\,$eV and
 $\alpha = -2.0$, E$ > 3\times 10^{15}\,$eV).
The factor involving energies 
in Eq.~\ref{eq:fract} accounts for the fact that
$\gamma$-ray primaries produce larger showers than cosmic-ray primaries
of the same energy, or conversely, that
$\gamma$-ray showers are detected
with the same efficiency as cosmic-ray showers of higher energy.
Therefore, to report a
flux ratio at a fixed {\em cosmic-ray energy}, we 
correct the $\gamma$-ray flux at its energy
to the higher cosmic-ray energy under the assumption that both species
have the same spectral index.
Table~\ref{tab:results} lists the 
values of $N_{90}$, $\epsilon_\gamma$, and
$I_\gamma / I_{cr}$ for each data sample.
We use measurements of $I_{cr}$ to
determine upper limits on the diffuse $\gamma$-ray flux 
at fixed {\em $\gamma$-ray energies,}
using the values of $N_{90}$, $N_{\rm tot}$, and $\epsilon_\gamma$.
The flux limits are given in Table~\ref{tab:results}.

\begin{table}[ht]
\caption{
Results of the search for diffuse ultrahigh energy $\gamma$-rays.
The median cosmic-ray energy, $E_{cr}$, and the median
gamma-ray energy, $E_\gamma$, are given in the first and fifth
columns, respectively, in units of TeV.
The quantities
$N_{90}$ and
$\epsilon_\gamma$ are defined in the text.
$I_\gamma/I_{cr}$ is the 
90\% C.L. upper limit on
the integral $\gamma$-ray fraction and
$I_\gamma$ is the 90\% C.L. upper limit on the
integral $\gamma$-ray flux, in
units of photons cm$^{-2}$ sec$^{-1}$ sr$^{-1}$.}
\label{tab:results}
\medskip
\begin{tabular}{cccccc}
$E_{cr}$  & $N_{90}$ & 
$\epsilon_\gamma$ &
$I_\gamma/I_{cr} (\times 10^{-5})$&
$E_{\gamma}$  &
$I_\gamma$ \\
\hline
575 &   7273.0 & 0.397 & $< 10.0$ &    330\ & 
        $< 1.0 \times 10^{-13}$ \\
1,350 &  1718.0 & 0.641 & $ < 6.5$ &    775\ & 
        $< 2.6 \times 10^{-14}$ \\
5,000 &  89.5  & 0.663 & $ < 4.0 $ &  2,450\ & 
        $< 2.1 \times 10^{-15}$ \\
22,000 & 2.3   & 0.736 & $ < 1.5 $ & 13,000\ & 
        $< 5.4 \times 10^{-17}$ \\
55,000 & 2.3   & 0.891 & $ < 8.0 $ & 33,000\ &  
        $< 5.3 \times 10^{-17}$ \\
\end{tabular}
\end{table}

\begin{figure}[ht]
\centerline{\psfig{file=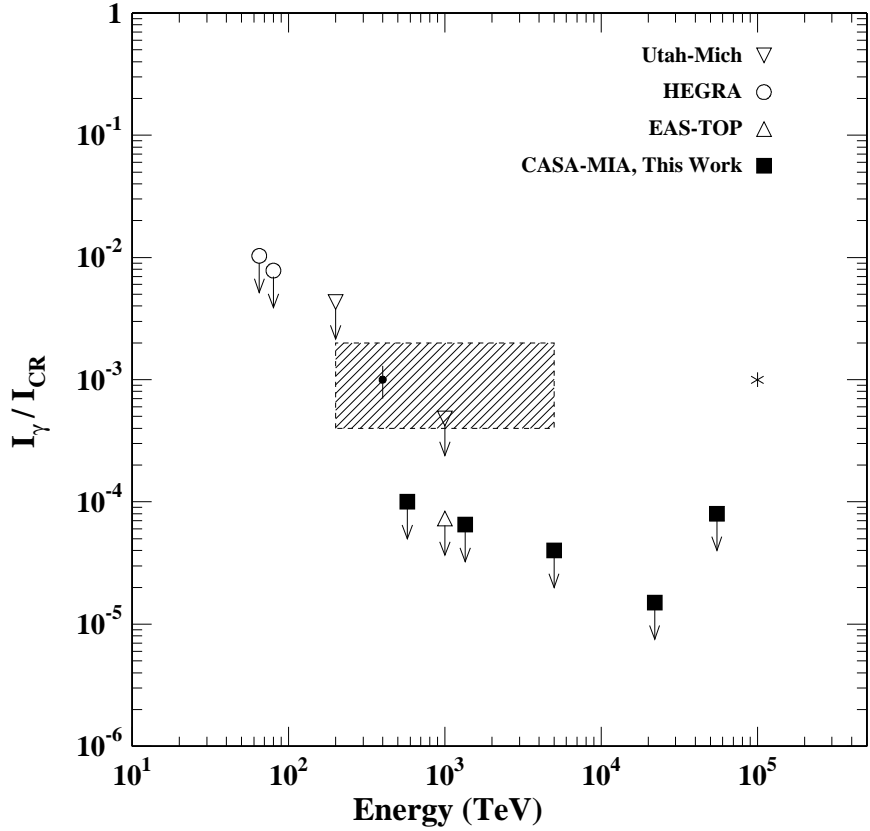,width=5.0in}}
\medskip
\caption[...]
{Measurements of the fraction of $\gamma$-rays relative to cosmic
rays at ultrahigh energies.
The hatched region indicates the range of $\gamma$-ray detections
as reported in the literature prior to 1985 [15,16].
The results of the
the Tien-Shan [17] and Yakutsk [18] experiments are shown by
the point with error bars and the asterix, respectively.
The points with arrows represent upper limits from
the Utah-Michigan [19], HEGRA [20], and
EAS-TOP [21] experiments, and this work, as indicated
in the legend.
\label{fig:limits}}
\end{figure}

Fig.~\ref{fig:limits} shows a compilation of measurements
on the $\gamma$-ray fraction as a function of energy, including our
work.
The limits presented here represent the first stringent results
spanning a wide range of energies,
including the first results above $2\times 10^{15}\,$eV.
The possibility of a diffuse $\gamma$-ray flux at the
$2\times 10^{-4}$ to $4\times 10^{-3}$ level, as suggested
by earlier results\cite{ref:lodz,ref:basje,ref:tienshan,ref:yakutsk}, 
is ruled out.
Our limits also constrain, to some extent, models
\cite{ref:td} which invoke topological defects as the cause of
cosmic radiation at the highest energies.
The limits disfavor models which have decaying topological defects
at relatively nearby distances (i.e. $< 100\,$Mpc), 
but not those which arrange
them in cosmologically uniform distributions \cite{ref:Protheroe}.

In summary, based on a very large sample of air shower events
whose muon content has been measured by a large muon detector,
we place stringent upper limits ($< 10^{-4}$)
on the $\gamma$-ray fraction of the cosmic rays
at energies between $5.7 \times 10^{14}$ and $5.5\times 10^{16}$.
These limits rule out the possibility of a significant 
component of the cosmic-ray flux whose interactions are
predominantly electromagnetic in nature (e.g. $\gamma$-rays,
electrons, positrons, etc.).
In addition, by using measurements of the cosmic-ray flux,
we place upper limits on the isotropic flux of $\gamma$-rays
at these energies,
which is of fundamental astrophysical interest.


\vspace{5mm}
\noindent {\bf Acknowledgements}
\vspace{3mm}

We acknowledge the assistance of the 
University of Utah Fly's Eye group and the
staff of Dugway Proving Ground.
We also thank P. Burke, S. Golwala, M. Galli, J. He, H. Kim, L. Nelson,
M. Oonk, M. Pritchard, P. Rauske, K. Riley, and Z. Wells 
for assistance with data processing.
Special thanks go to M. Cassidy.
This work is supported by the U.S. 
National Science Foundation and the U.S. Department of Energy.
JWC and RAO acknowledge the support of
the Grainger Foundation and
the Alfred P. Sloan Foundation.

\vspace{10mm}

\noindent $^\dag$ Present Address: Department of Physics and Astronomy,
Iowa State University, Ames, IA 50011, USA.

\newpage


\end{document}